\documentclass[journal]{IEEEtran}
\ifCLASSINFOpdf

\else

\fi

\usepackage[cmex10]{amsmath}
\usepackage{amsfonts}
\usepackage{array,color}
\usepackage{booktabs}
\usepackage{graphicx}
\usepackage{epstopdf}
\usepackage{cite}
\usepackage{algorithm}
\usepackage{algorithmic}
\usepackage{amssymb}
\usepackage{stfloats}
\begin{document}

\title{Towards SE and EE in 5G with NOMA and Massive MIMO Technologies}
\author{
\IEEEauthorblockN{Di Zhang\IEEEauthorrefmark{1}, Zhenyu Zhou\IEEEauthorrefmark{2}, and Takuro Sato\IEEEauthorrefmark{1}}

\IEEEauthorblockA{\IEEEauthorrefmark{1} GITS/GITI, Waseda University, Tokyo, Japan, 169--8050\\
Email: di\_zhang@fuji.waseda.jp, t-sato@waseda.jp\\}
\IEEEauthorblockA{\IEEEauthorrefmark{2} National Key Laboratory of Alternate Electrical Power System with Renewable Energy Sources,\\
School of Electrical and Electronic Engineering, North China Electric Power University, Beijing, China, 102206\\
Email: zhenyu\_zhou@fuji.waseda.jp}
}

\maketitle

\begin{abstract}
Non-Orthogonal Multiple Access (NOMA) has been proposed to enhance the Spectrum Efficiency (SE) and cell-edge capacity. This paper considers the massive Multi-Input Multi-Output (MIMO) with Non-Orthogonal Multiple Access (NOMA) encoding. The close-form expression of capacity of the massive MIMO with NOMA is given here. Apart from the previous Successive Interference Cancellation (SIC) method, the Power Hard Limiter (PHD) is introduced here for better reality implement.  
\end{abstract}


\IEEEpeerreviewmaketitle

\section{Introduction}
Within the background of 5G research, the EE issue was proposed as another important elements together with the previous SE issue. Nowadays, while studying about the SE and EE, some potential techniques were proposed, such as the massive MIMO\cite{1}, mmWave\cite{2}, NOMA\cite{3}, etc. For mmWave, the main work was waiting for further spectrum resources allocation from ITU. For massive MIMO, some prior works had been done based on the existing MIMO technologies, such as the cellular zooming\cite{4}. And thanked to this proposal, once proposed,a great deal of studies erupted in EE studies. For instance, in\cite{5} \cite{6}, some power allocation mechanisms were proposed towards massive MIMO, but antenna selection was focused in those paper whereas the BS coverage area was neglected. Although lots of studies had been done, but we noticed that if we combine the NOMA together with massive MIMO, by some potential techniques, we can further enhance the EE performance. Which relied in the fact that energy consumption and achievable total transmission rate by a certain bandwidth are two factor of EE performance.
\section{Analysis of Massive MIMO system with NOMA}
\subsection{Capacity Analysis of NOMA with Signal Antenna}
Consider that in a massive MIMO system, there are $M$ antennas serving for a number of $K$ user equipments (UEs). While the signal information transmitted from the antennas to the UEs, the received $K \times 1$ signals can be demonstrated as:
\begin{equation}
\textbf{y} = \sqrt{P_k} \textbf{Hx} + \textbf{n},
\end{equation}
where \textbf{H} represents for the $M \times K$ channel matrix between the transmit antennas and received UEs. $P_k$ denotes for the power of each signal that transmitted from antenna to received UEs whereas $\textbf{x}$, $\textbf{n}$ represent for the transmitted signal and additive white gaussian noises (AWGN), where the AWGN is a $\textbf{n} \sim \mathcal{C}(0, 1)$ additive noises. Here $\sqrt{P_k} \textbf{x}$ is the simultaneously transmitted symbol vector of the signals.
\par
The channel matrix models the independent fast fading, slow fading which is also can be expressed by the geometric attenuation and log-normal shadow fading. Thus the channel coefficient $h_{m,k}$ can be demonstrated as $h_{m,k} = a_{m,k} \sqrt{\beta_{m,k}}$ where $a_{m,k}$ is the fast fading coefficient from the $m$th antenna to $k$th UE. $\sqrt{\beta_{m,k}}$ models the geometric attenuation and shadowing fading coefficient, which can be demonstrated by the distance factor while ignoring the shadowing fading influence and can be expressed as
$\beta_{m,k} = \frac{1}{1+d_{m,k}^\alpha}$. Here the $d_{m,k}$ denotes the distance from antenna to the received UE with $\alpha$ as the path loss factor. Without loss of generality, we suppose that in one coverage area, the BS is located in the center with radius $\mathcal{R_D}$ where UEs uniformly distributed within the coverage area.
\par
Take the power coefficient as $\gamma$, while adopting the non-orthogonal multiple access (NOMA) and successive interference cancellation (SIC), which will successively remove the signal of other UEs while $i < k$ and treat the $i > k$ as the noise, thus we can get the achievable data rate of $k$th UE as:
\begin{equation}
R_k = \log_2(1 + \frac{\rho |h_{m,k}|^2 \gamma_k}{\rho |h_{m,k}|^2 \sum_{i = k+1}^{K}\gamma_k+1} ),
\end{equation}
where $\rho$ represents for the transmit SNR. Be aware that the data rate of UE with number $K$ is $R_K = \log_2 (1 + \rho |h_{K}|^2 \gamma_K)$. Suppose that the channel information will determine the data rate of each UE, taking $\tilde{\gamma_k} = \sum_{i = m+1}^{M} \gamma_{k}$, the sum rate achieved by NOMA can be expressed by:
\begin{equation}
R_{sum} = \sum_{m = 1}^{M-1} \log_2(1 + \frac{\rho |h_{m,k}|^2 \gamma_k}{\rho |h_{m,k}|^2 \tilde{\gamma_k}+1}) + \log_2 (1 + \rho |h_{K}|^2 \gamma_K),
\end{equation}
thus suppose that the distance between UEs and antennas are much larger than the distance between the antennas, in addition, on condition that the data rate request of each UE is fulfilled. The sum ergodic rate is given by:
\begin{equation}
\begin{aligned}
R_{ave} &= \sum_{m = 1}^{M-1} \int_{0}^{\infty} \log_2(1 + \frac{x \rho \gamma_k}{ x \rho \tilde{\gamma_k}+1})f_{|h_{m,k}|^2}(x)dx \\
&+ \int_{0}^{\infty}  \log_2(1 + x\rho \gamma_K)f_{|h_{K}|^2}(x)dx,
\end{aligned}
\end{equation}
where $f_{|h_{m,k}|^2}, f_{|h_{K}|^2}$ represent for the probability density function (PDF) of the channel gain, which can be given as\cite{8}:
\begin{equation}
f_{|h_{m,k}|^2}(x) \approx \frac{1}{\mathcal{R_D}}\sum_{j = 1}^{M}\sum_{k = 1}^{K}\delta_{m,k} e^{-c_{m,k}x},
\end{equation}
where $\delta_{m,k} = \omega_{m,k}\sqrt{1 - \theta_{m,k}^{2}}(\frac{\mathcal{R_D}}{2}\theta_{m,k} + \frac{\mathcal{R_D}}{2})c_{m,k}$ with $\theta_{m,k} = cos(\frac{2k-1}{2K}\pi)$ and $c_{m,k} = 1 + (\frac{\mathcal{R_D}}{2}\theta_{m,k} + \frac{\mathcal{R_D}}{2})^{\gamma_{m,k}}$. This can be obtained by applying the Gaussian-Chebyshev quadrature (GCQ).
\par
According to the deduction in \cite{7}, the achievable ergodic sum rate of NOMA can be given as:
\begin{equation}
R_{ave} \to \log_2 (\rho \log_2 \log_2 K).
\end{equation}
\subsection{Capacity Analysis of NOMA with massive MIMO}
For massive MIMO system with traditional encoding methods, it is proved that Minimum Mean-Square Error Detector (MMSE) can reach the maximum achievable rates in uplink with perfect CSI\cite{8}, whereas maximum achievable downlink rates gained by MMSE or Regularized Zero-Forcing (RZF)\cite{9}

\section{Energy Efficiency Analysis of Massive MIMO with NOMA}
While solving the antenna and radio frequency (RF) chain selection problems, suppose that in each BS coverage area, the distribution of UEs obeys the same PPP distribution and the requirements of capacity are also following the same Poisson distribution, the optimal solution can be gained by solving the problem in one BS area of one epoch time. Thus the optimal antenna and RF chain selection problem can be written as:
\begin{equation}
\begin{aligned}
\label{eq:14}
\underset{P_k,  P_{RF}}{\text{max}}\quad \frac{R_{ave}}{P_k + P_{RF}},  
\end{aligned}
\end{equation}

\begin{subequations}
\label{eq:14con}
\begin{align}
    \text{subject to:} \nonumber\\
         C1: &\sum_{k=1}^{K} (\frac{1}{\eta}P_k + P_c + P_{RF}) \leq P_{T}, \label{eq:14con1}\\
         C2: &\sum_{k=1}^{K}  N_{k,b}^{a} \leq N_{bs}^{a}, \forall b,  \label{eq:14con2} \\
         C3: &N_{k,c}^{UE} \leq   \sum_{k=1}^{K}  N_{k,c}^{rf} \leq N_{bs}^{rf}, \forall t, \label{eq:14con3} \\
         C4: &P_k, P_c, P_{RF} \geq 0, \forall k. \label{eq:14con4}
\end{align}
\end{subequations}
\par
If we define a maximum weighted solution by $S^* = \text{max}(P_k,  P_{RF})= \sum_{k=1}^{K} R_{k}^*/(P_k + P_{RF})^*$ the problem has feasible solutions and its optimum solution is $S^*$, then the existing optimal solution if any, can only achieved by:
\begin{equation}
\begin{aligned}
&\text{max}(P_k,  P_{RF}) - S^* (P_k + P_{RF})^* \\
& = \sum_{k=1}^{K} R_{k}^* - S^* (P_k + P_{RF})^* = 0,
\end{aligned}
\end{equation}
and the optimal solution of $\text{max}(P_k,  P_{RF})$ should be equal or smaller than $S^*$ if any. Note that the RF chain power consumption has nothing related with the achievable capacity of one cluster although it is needed for transmission. In this case, while searching for the optimal solver, it will converge to zero. Here as the general assumption, we assume that in each step, the number of RF chain is equal to the number of antenna in order to serving the requirement of each UE. Thus the extreme point can be verified by the lagrange method:
\begin{equation}
\begin{aligned}
&\frac{\partial \text{max}(P_k,  P_{RF})}{\partial P_k} = \frac{\partial \frac{R_{ave}}{P_k + P_{RF}}}{\partial P_k} \\
&= \frac{\frac{1}{P_k \ln 2} - \log_2(\frac{P_k}{N_0}\log_2\log_2K)}{(P_k + P_{RF})^2},
\end{aligned}
\end{equation}

\begin{equation}
\begin{aligned}
&\frac{\partial^2 \text{max}(P_k,  P_{RF})}{\partial (P_k)^2} = \frac{\partial^2 \frac{R_{ave}}{P_k + P_{RF}}}{\partial (P_k)^2} \\
&= \frac{\frac{-\ln2}{(P_k\ln2)^2}-\frac{1}{P_k\log_2\log_2K\ln2}}{(P_k + P_{RF})^2} \\
&- \frac{2[\frac{1}{P_k\ln2}-\log_2(\frac{P_k}{N_0}\log_2\log_2K)]}{(P_k + P_{RF})^3}.
\end{aligned}
\end{equation}
On condition that high SNR with limited number of UEs in one cluster, we can conclude that the second derivative should be less than zero. Thus it is easily to know that the equation of power consumption $P_k + P_{RF}$ is a affine function of $P_k$, and the objective is a quasi-convex function. Based on our deductions, this can be solved by the subgradient method, which for a given accuracy, it will search for the optimal point that meets the requirement of accuracy.
\section{conclusion}
In this paper, the massive MIMO with NOMA's achievable ergodic sum rate is given. In addition, the EE performance is analyzed here. In the following step, we will further our study while comparing with the previous studies and conclude the proposal.
\section*{acknowledgement}
The author would like to thank Chinese Scholarship Council (CSC) for its financial support of this study under grant 201306770001.
\begin{bibliographystyle}{IEEEtran}
\begin{bibliography}{IEEEabrv,thirdpaper}
\end{bibliography}
\end{bibliographystyle}
\end{document}